\documentclass[twocolumn,aps,pra,superscriptaddress,10pt]{revtex4-2}
\usepackage{graphicx}
\usepackage{amsmath}
\usepackage{amssymb}
\usepackage{braket}
\usepackage{bm}
\usepackage{dcolumn}
\usepackage{mathrsfs}
\usepackage{subfigure}
\usepackage{graphicx}
\usepackage[colorlinks=true,linkcolor=blue,citecolor=blue,urlcolor=blue]{hyperref}
\usepackage{soul}
\usepackage{changes}
\usepackage{amsthm,amsmath,amssymb}
\usepackage{float}
\usepackage{mathrsfs}
\usepackage[toc,page,title,titletoc,header]{appendix}
\usepackage[utf8]{inputenc}
\usepackage[T1]{fontenc}
\makeatother
\begin{document}

\title{Eikonal Approximation for Floquet Scattering}

\author{Yaru Liu}

\affiliation{Department of Physics, Renmin University of China, Beijing, 100872,
China}

\affiliation{Key Laboratory of Quantum State Construction and Manipulation (Ministry of Education), Renmin University of China, Beijing, 100872, China}

\author{Peng Zhang}
\email{pengzhang@ruc.edu.cn}
%\selectlanguage{english}%
\affiliation{Department of Physics, Renmin University of China, Beijing, 100872,
China}

\affiliation{Key Laboratory of Quantum State Construction and Manipulation (Ministry of Education), Renmin University of China, Beijing, 100872, China}

\begin{abstract}
The eikonal approximation (EA) is widely used in various high-energy scattering problems. In this work we generalize this approximation from the scattering problems with time-independent Hamiltonian to the ones with periodical Hamiltonians, {\it i.e.}, the Floquet scattering problems. We further illustrate the applicability of our  generalized EA  via the scattering problem with respect to a shaking spherical square-well potential,
 by comparing the results given by this approximation and the exact ones. The generalized EA  we developed is helpful for the research of manipulation of high-energy scattering processes with external field, {\it e.g.}, the manipulation of atom, molecule or nuclear collisions or reactions via strong laser fields. 

 \vspace{0.5cm}
 \noindent {\color{black}\textbf{Keywords:} Eikonal Approximation, Floquet Scattering, high-energy scattering.} 
\end{abstract}
\maketitle

% \noindent {\color{black}\textbf{Keywords:} Eikonal Approximation, Floquet Scattering, high-energy scattering, periodical systems}

\section{ Introduction}

The eikonal approximation (EA) for quantum scattering was developed by R. Glauber\cite{glauber1987high}. As one of the most widely used
approximation of quantum scattering problems, the EA was applied 
 in numerous  high-energy scattering problems of various physical systems, such as nuclear collisions\cite{wallace1975high,obu1972generalized,schiff1968high,frahn1974high,hebborn2017analysis}, nuclear reactions\cite{bianconi1995test,debruyne2000relativistic,buuck2014corrections}, or the collisions between other types of high-energy particles\cite{aguiar1997low,goldhaber1968high,murphy1975eikonal,meggiolaro1998high,fukui2014analysis,wilets1968eikonal,sargsian2001selected,esbensen2001eikonal}. The Hamiltonians of these scattering problems are given by the relevant natural interactions, and thus are
  time-independent.
  
  On the other hand, in recent years 
  the technique of manipulation of scattering processes with  laser or other kinds of periodical external fields has been developed for various physical systems\cite{otten1981laser,hannachi2007prospects,negoita2022laser,gales2016new,ur2015eli,matinyan1998lasers,wang2022nuclear,wang2021exciting,qi2020nuclear,wang2022strong,xu2023laser,xu2009optically,qi2023isomeric}. In the presence of the periodical field, the Hamiltonian of the scattering problem becomes a time-dependent one. These problems should be treated by the Floquet scattering theory\cite{moskalets2002floquet,li1999floquet,bilitewski2015scattering,emmanouilidou2002floquet}, and
 the "original form" of the EA, which is developed for the scattering with time-independent Hamiltonian, cannot be directly applied. 
  
  In this work we generalize the EA to the scattering problems with  periodical Hamiltonians. We derive the expressions of the scattering wave function and scattering amplitude  given by the EA, and verify our results via the example of shaking spherical square well potential.

  Notice that the original EA was initially developed for the one-body potential scattering problem, and was later applied in more complicated scattering problems, {\it e.g.}, the nuclear reaction problems\cite{wallace1975high,obu1972generalized,schiff1968high,frahn1974high,hebborn2017analysis,bianconi1995test,debruyne2000relativistic,buuck2014corrections}. Similarly, 
although in this work we focus on the one-body potential scattering, which is for simplicity, the generalized EA  we developed can also be applied for more complicated scattering problems with periodical Hamiltonian.  
  
The remainder of this paper is organized as follows. In Sec.~\ref{swf}
we derive the generalized
EA   for the calculations of the 
Floquet scattering wave function, scattering amplitude and 
 and cross section. In Sec.~\ref{sw} we apply the generalized EA for a periodical spherical square-well model, and show the applicability of our approach.
There is a summary in Sec.~\ref{summary}.
Some details of our calculation approach are shown in the appendixes.

\section{Generalized EA}
\label{swf}

In this section we generalize the EA for the Floquet scattering. We first derive the EA for the calculation of Floquet scattering wave function, and then  calculate the Floquet scattering amplitude and cross section with the wave function obtained by the EA.

\subsection{Scattering Wave Function}

We consider the scattering of a single particle on a periodical potential, with the Hamiltonian being given by
\begin{eqnarray}
    \hat{H}=-\frac{\hbar^2}{2m}\nabla^2+U({\bm r},t).\label{h}
\end{eqnarray}
Here $m$ and ${\bm r}\equiv x{\bm e}_x+y{\bm e}_y+z{\bm e}_z$ 
 are the mass and position of the particle, respectively,  with ${\bm e}_{x,y,z}$ being the unit vector along the $x,y,z$ directions.
In addition, $U({\bm r},t)$ is the  scattering potential, which is a periodical function of time and satisfies
\begin{eqnarray}
      U({\bm r},t)&=&U\left({\bm r},t+T \right),\label{uc}
\end{eqnarray}
with $T$  being the period, and $
\lim_{r\rightarrow\infty}U({\bm r},t)=0$.

We consider the scattering process with incident momentum $\hbar{\bm k}$. Without loss of generality, in this work we assume ${\bm k}$ is along the $z$-direction, {\it i.e.}, 
\begin{eqnarray}
{\bm k}=k{\bm e}_z.
\end{eqnarray}
According to the Floquet scattering theory, the corresponding scattering wave function $\Psi({\bm r},t)$ of our system satisfies the time-dependent  
Schr\"odinger equation
\begin{eqnarray}
    {\color{black}\mathrm{i}}\hbar\frac{\partial}{\partial t}\Psi({\bm r},t)={\hat H}\Psi({\bm r},t),\label{se}
\end{eqnarray}
and can be written as
\begin{eqnarray}\label{wavefunction_tot}
     \Psi({\bm r},t)={\color{black}\mathrm{e}}^{-{\color{black}\mathrm{i}}Et/\hbar}\psi({\bm r},t).\label{wf}
 \end{eqnarray}
Here $\psi({\bm r},t)$ is a periodic function satisfying 
\begin{eqnarray}
    \psi({\bm r},t)=\psi({\bm r},t+T),\label{psip}
 \end{eqnarray}
and
$E\equiv \hbar^2k^2/(2m)$ is the scattering energy, with $k=|{\bm k}|$.

Now the question is how to derive the periodic function $\psi({\bm r},t)$. 
As in the derivations for the EA for time-independent potential\cite{landau2013quantum}, here we consider the high-energy cases with 
\begin{eqnarray}
k\gg 2\pi/l_U,\ \ \ E\gg U_\ast,\label{con}
\end{eqnarray}
 where $l_U$ and $U_\ast$ are the characteristic length scale and characteristic strength of the potential $U({\bm r},t)$, respectively.
Under the condition (\ref{con}), the scattering effect is weak, so that the wave function $\psi({\bm r},t)$ can be approximated as the product of the incident wave and a factor which is slowly varying in the spatial space, {\it i.e.},
 \begin{eqnarray}\label{wavefunction_tot}
     \psi(\bm{r},t)\approx\frac{1}{(2\pi)^{3/2}}{\color{black}\mathrm{e}}^{{\color{black}\mathrm{i}}kz}\phi_{\rm EA}(\bm{r},t),\label{psirt}
 \end{eqnarray}
 where $\phi_{\rm EA}(\bm{r},t)$ is a slowly-varying function of ${\bm r}$ 
% {\color{black}[add a footnote: In the Floquet scattering, in principle, there exists some channels with  outgoing momentum being much less than $k$ ({\it i.e.}, the channels with $n$ being close to $n_\ast$, where $n$ and $n_\ast$ being defined in Sec.~\ref{sacs}). The wave function of these channels
% cannot be expressed as $a slowly-varying function of ${\bm r}$$
% ]}. 
 In addition, due to the condition (\ref{psip}),  $\phi_{\rm EA}(\bm{r},t)$ is a
a periodic function of $t$  {\it i.e.},
 \begin{eqnarray}
    \phi_{\rm EA}({\bm r},t)=\phi_{\rm EA}({\bm r},t+T).\label{ppc}
 \end{eqnarray}
Furthermore,  substituting Eqs.~(\ref{wf},\ref{wavefunction_tot}) into the Schrodinger equation \eqref{se}, we obtain the explicit equation satisfied by $\phi_{\rm EA}({\bm r},t)$:                                  
\begin{eqnarray}\label{unreduce_eq}
   \left[{\color{black}\mathrm{i}}\hbar\frac{\partial }{\partial t} +
   \frac{{\color{black}\mathrm{i}}\hbar^2k}{2m} \frac{\partial }{\partial z} +\frac{\hbar^2}{2m}\nabla ^2-U({\bm r},t) \right]\phi_{\rm EA}({\bm r},t)=0.\label{eeq}
\end{eqnarray}
{\color{black}
Notice that $\phi_{\rm EA}({\bm r},t)$ is a slowly varying function of ${\bm r}$. Explicitly,  
the characteristic length scale $l_\ast$ of the spatial variation of $\phi_{\rm EA}({\bm r},t)$ satisfies $1/l_\ast \gg k$. On the other hand, 
$\partial\phi_{\rm EA}({\bm r},t)/\partial z$  and
$\nabla^2 \phi_{\rm EA}({\bm r},t)$ are on the order of $\phi_{\rm EA}({\bm r},t)/l_\ast$ and $\phi_{\rm EA}({\bm r},t)/l_\ast^2$, respectively. As a result, in Eq.(10) the term
$(\frac{\hbar^2}{2m})\nabla^2 \phi_{\rm EA}({\bm r},t)$ is much less than $(\frac{\hbar^2}{2m})k\frac{\partial\phi_{\rm EA}({\bm r},t)}{\partial z}$, and thus can be ignored\cite{1page}. Then we obtain}
\begin{eqnarray}\label{DIF}
   \left[ {\color{black}\mathrm{i}}\frac{\partial }{\partial t}+{\color{black}\mathrm{i}} v_z \frac{\partial }{\partial z}-\frac{1}{\hbar}U({\bm r},t)\right]\phi_{\rm EA}({\bm r},t)=0,\label{appe}
\end{eqnarray}
where $v_z=\hbar k/m$ is the velocity of the incident particle. 

For the convenience of the following discussion, we define the vector ${\bm b}$ as the projection of the position ${\bm r}$ on the $x$-$y$ plane, {\it i.e.},
\begin{eqnarray}
{\bm b}=x{\bm e}_x+y{\bm e}_y.
\end{eqnarray}
Thus, the potential $U({\bm r},t)$ is also function of $\{{\bm b},z,t\}$, {\it i.e.}, 
\begin{eqnarray}
U({\bm r},t)=U({\bm b},z,t).
\end{eqnarray}
We find that with the above notations, the solution of Eq.~(\ref{appe}) can be expressed as
\begin{eqnarray}
\phi_{\rm EA}({\bm r},t)=\exp\left\{ -{\color{black}\mathrm{i}}\frac{1}{v_z\hbar}\int_{-\infty} ^{z}
U\bigg[
{\bm b}, z', {\tilde t}(t,z,z')
\bigg]{\color{black}\mathrm{d}}z'
     \right\},  \nonumber\\   \label{cr}
\end{eqnarray}
where
\begin{eqnarray}
   {\tilde t}(t,z,z')&=&t+\frac{z'-z}{v_z}.\label{cr2}
\end{eqnarray}
One can verify this solution  by directly substituting Eqs.~(\ref{cr}-\ref{cr2}) into Eq.~(\ref{appe}). Additionally,
$\phi_{\rm EA}({\bm r},t)$  tends to $1$ in the limit $z\rightarrow-\infty$ for arbitrary $t$, 
as the slowly-varying wave function in the  EA for time-independent potential,
and satisfies the periodical condition (\ref{ppc}). The latter fact can be directly proved via the fact $U({\bm r},t)=U\left({\bm r},t+T \right)$. Moreover, when $U$ is independent of $t$, Eq.~(\ref{cr}) becomes
$\phi_{\rm EA}=\exp\{ -{\color{black}\mathrm{i}}\frac{1}{v_z\hbar}\int_{-\infty} ^{z}
U[
{\bm b}, z']{\color{black}\mathrm{d}}z'\}$. This is just the slowly-varying part of the scattering wave function for time-independent potential, which is given by the traditional EA\cite{landau2013quantum}.

In summary,  the scattering wave function  
given by the generalized  EA  for the Floquet scattering is 
$\Psi({\bm r},t)=(2\pi)^{-3/2}{\color{black}\mathrm{e}}^{-{\color{black}\mathrm{i}}Et/\hbar}{\color{black}\mathrm{e}}^{{\color{black}\mathrm{i}}kz}\phi_{\rm EA}(\bm{r},t)$, with the function $\phi_{\rm EA}(\bm{r},t)$ being given by Eq.~(\ref{cr}).{\color{black}\cite{2page}}

%given by Eqs.~(\ref{wf}, \ref{psirt}, \ref{cr}).

\subsection{Scattering Amplitude}
\label{sacs}

Now we calculate the scattering amplitude  via the Floquet scattering wave function $\Psi({\bm r},t)$ 
obtained above via the generalized EA.

%\subsection{Scattering Amplitude}

For our scattering problem, since the potential $U$ is time-dependent, the  energy conservation is broken.
As a result, for 
given
 incident momentum ${\bm k}$, the kinetic energy of the particle   after the scattering process can 
 be
 \begin{eqnarray}
 E_n\equiv \frac{\hbar^2k^2}{2m}+n\hbar\omega,\ \ \ {\rm with}\ n=n_{\ast}, n_{\ast}+1,...,\nonumber\\
 \end{eqnarray}
 where 
 \begin{eqnarray}
 \omega=2\pi/T,
 \end{eqnarray}
 and $n_{\ast}\leq 0$ is the minimum integer which satisfies $E+n\hbar\omega\geq 0$.
Furthermore, according to the Floquet scattering theory (Appendix~\ref{fsa}), the scattering amplitude with respect to incident momentum ${\bm k}=\hbar k{\bm e}_z$ and outgoing momentum ${\bm k}'$, which satisfy 
\begin{eqnarray}
 \frac{\hbar^2|{\bm k}'|^2}{2m}=E_n,\ \ \ (n=n_{\ast}, n_{\ast}+1,...),
\end{eqnarray}
is given by (Appendix~\ref{fsa})
% \begin{widetext}
\begin{eqnarray}
 &&f({\bm k}', n\leftarrow {\bm k})\nonumber\\
 &=& -\frac{m\omega}{(2\pi)^{1/2}}\int_{0}^{2\pi/\omega}\!\!{\color{black}\mathrm{d}}t 
    \int_{-\infty}^{+\infty} \!\!{\color{black}\mathrm{d}}\bm r  {\color{black}\mathrm{e}}^{-{\color{black}\mathrm{i}}{\bm k'}\cdot {\bm r}} {\color{black}\mathrm{e}}^{{\color{black}\mathrm{i}}n\omega t/\hbar}U(\bm r,t)\psi(\bm r,t),
    \nonumber\\
    \label{fkkp}
\end{eqnarray}
 where $\psi({\bm r},t)$ is related to the scattering wave function $\Psi(\bm r,t)$ via Eq.~(\ref{wf}). Substituting Eq.~(\ref{psirt})  into Eq.~(\ref{fkkp}), we find that under the generalized EA, the Floquet scattering amplitude  can be expressed as
 \begin{eqnarray}
 &&f({\bm k}',n\leftarrow {\bm k})\nonumber\\
& \approx&  -\frac{m\omega}{(2\pi)^{2}}
\int_{0}^{T} \!\!\!{\color{black}\mathrm{d}}t 
    \!\int_{-\infty}^{+\infty} \!\!\! {\color{black}\mathrm{d}}\bm r  {\color{black}\mathrm{e}}^{-{\color{black}\mathrm{i}}({\bm k'}-{\bm k})\cdot {\bm r}} {\color{black}\mathrm{e}}^{{\color{black}\mathrm{i}}n\omega t}U(\bm r,t)\phi_{\rm EA}(\bm r,t),\nonumber\\ \label{fea}
 \end{eqnarray}
 with the wave function $\phi_{\rm EA}(\bm r,t)$ being given by Eq.~(\ref{cr}).
%  \begin{figure}[tbp]
%     \includegraphics[width=0.7\columnwidth]{FIG/sigtheta2.pdf}
%     \caption{The total scattering cross section is a function of $\theta$,$U_0=U_1=10$.}
%     \label{theta}
% \end{figure} 

 Moreover, under the high-energy condition (\ref{con}) of the EA, these scattering amplitudes
 with the 
 outgoing momentum $\hbar{\bm k}'$ satisfying 
 \begin{eqnarray}
 |{\bm k}'|=k\ \ {\rm and}\ \  |{\bm k}'-{\bm k}|\ll k,\label{ccon}
 \end{eqnarray}
  ({\it i.e.}, 
  $n=0$ and
  the angle between the incident and outgoing momentum is small) are much larger than the  ones 
  for the outgoing momentums not satisfying Eq. (\ref{ccon}).
  As proven in Appendix ~\ref{pro1}, under the condition (\ref{ccon}), the scattering amplitude $f({\bm k}',0\leftarrow {\bm k})$ given by the EA ({\it i.e.}, Eq.~(\ref{fea})) can be further reduced to 
 \begin{eqnarray}
&&f({\bm k}',0\leftarrow {\bm k})\nonumber\\
&\approx& \frac{\omega  \hbar^2k}{(2 \pi)^{2}{\color{black}\mathrm{i}}}\int_{0}^{T}\,{\color{black}\mathrm{d}}t\!\int_{-\infty}^{+\infty} \!\!\! {\color{black}\mathrm{d}}\bm b {\color{black}\mathrm{e}}^{-{\color{black}\mathrm{i}}({\bm k'}-{\bm k})\cdot {\bm b}}  \nonumber\\
&&  \left\{\exp\left( -{\color{black}\mathrm{i}}\frac{1}{v_z\hbar}\int_{-\infty} ^{+\infty}
U\bigg[
{\bm b}, z', {\tilde t}(t,z,z')
\bigg]{\color{black}\mathrm{d}}z'
     \right)-1\right\}\ \ \ \ \ \ \nonumber\\
     \nonumber\\
  &&   ({\rm for}\  k=k',\ \  {\rm and}\ \ |{\bm k}'-{\bm k}|\ll k).
  \nonumber\\
  \label{fea2}
\end{eqnarray}
Moreover,
when the potential $U({\bm r},t)\equiv U({\bm b},z,t)$ is independent of the direction of ${\bm b}$, {\it i.e.}, $U({\bm b},z,t)\equiv U(b,z,t)$,  Eq.~(\ref{fea2}) can be further approximated as (Appendix ~\ref{pro1}):
\begin{eqnarray}
\label{ftheta}
       &&f({\bm k}',0\leftarrow {\bm k})\nonumber\\
&\approx&\frac{ \omega \hbar^2 k}{2 \pi {\color{black}\mathrm{i}}}\int_{0}^{T}\,{\color{black}\mathrm{d}}t\!\int_{0}^{+\infty} b\,{\color{black}\mathrm{d}}b J_0(k b \theta)
\nonumber\\
        &&\left\{\exp\left( -{\color{black}\mathrm{i}}\frac{1}{v_z\hbar}\int_{-\infty} ^{+\infty}
    U\bigg[
     b, z', {\tilde t}(t,z,z'),
    \bigg]{\color{black}\mathrm{d}}z'
         \right)-1\right\},\nonumber\\
         \label{ff0}
\end{eqnarray}
where $\theta$ is the angle between ${\bm k}$
and ${\bm k}^\prime$, and $J_0$ is the 0th order Bessel function of the first kind.

   %\end{widetext}
   
\subsection{Crosssection}
   
  Using the scattering amplitude $ f({\bm k}',n\leftarrow {\bm k})$ given by Eq.~(\ref{fea}), one can further derive various cross section of the scattering process. In particular, the differential cross section with respect to outgoing kinetic energy $E_n$ and outgoing momentum direction ${\hat {\bm s}}$ is 
 \begin{eqnarray}
 \frac{d\sigma(E_n,{\hat{\bm s}})}{d{\hat{\bm s}}}=\sqrt{\frac {k_n}k}\big\vert f(k_n{\hat{\bm s}},n\leftarrow {\bm k})\big\vert^2,
 \end{eqnarray}
with $k_n=\sqrt{2mE_n/\hbar}$. Furthermore, according to the optical theorem, the total cross section with respect to incident momentum ${\bm k}$ is
\begin{equation}
    \sigma_{\rm tot}(k)=\frac{4\pi}{k}{\rm Im}\big[f(\bm k,0 \leftarrow \bm k)\big].
    \label{opt}
\end{equation}

%{\color{blue}Next, we discuss the variations of and  scattering amplitude $f$ and scattering cross section $\sigma_{\rm tot}$ with scattering angle $\theta$ 
%%\cite{glauber1987high}
%
%According to the Kramers-Kronig relations, for a given incident momentum $k$, the scattering cross-section in the $\theta$ direction can be expressed as:
%\begin{equation}
%    \sigma(\theta)=\frac{4\pi}{k^2}|f(\theta)|^2.
%    \label{kkr}
%\end{equation}
%}

\begin{figure*}[tbp]
    \includegraphics[width=1.8\columnwidth]{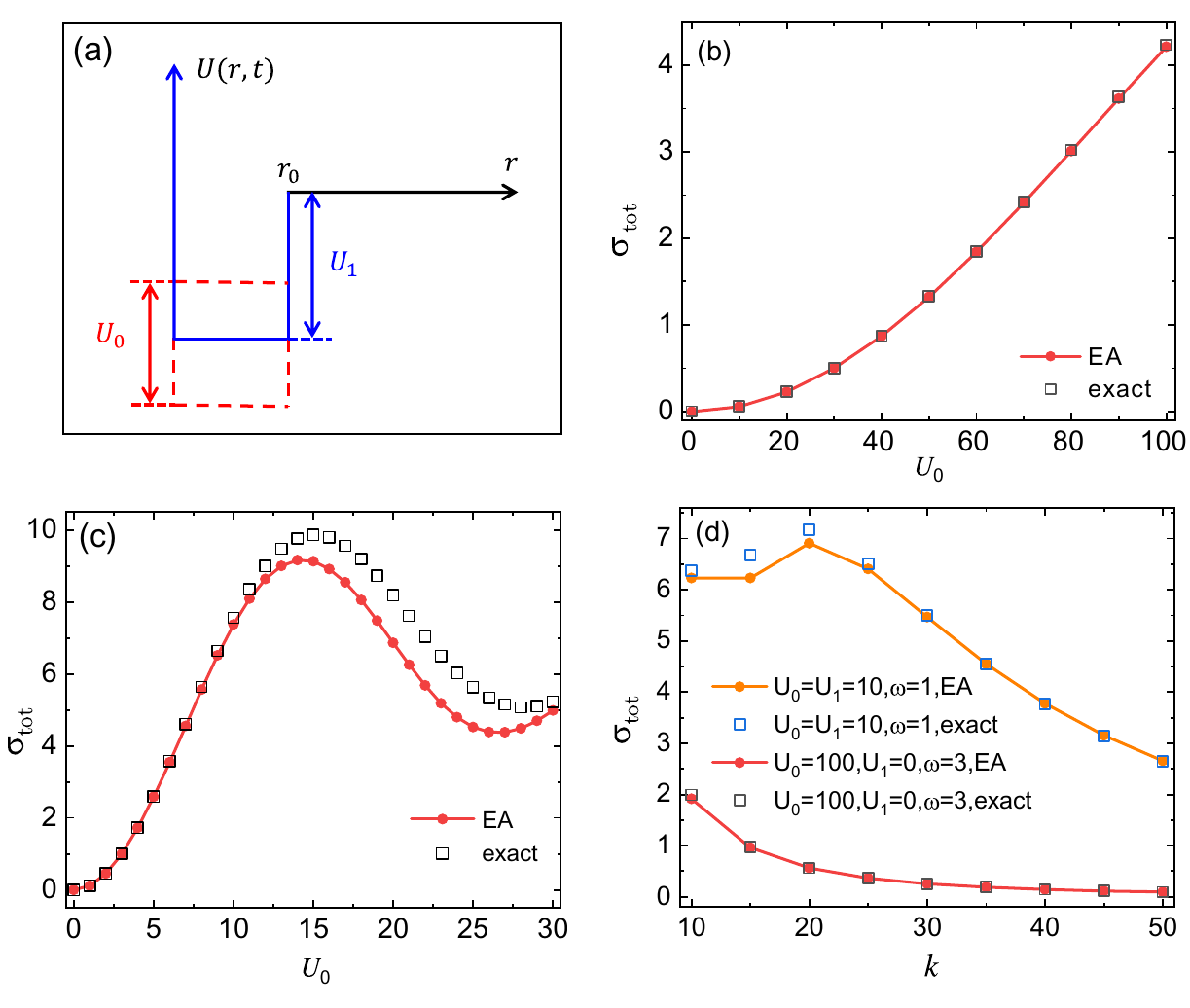}
    \caption{
 {\bf (a):} Schematic diagram of  the spherical square-well with shaking depth. {\bf (b-d):}
    Total scattering cross section $\sigma_{\rm tot}(k)$ of the shaking square-well model.  Here we show the results given by the exact numerical calculations (squares) and the generalized EA (dots connected by solid lines) we developed in this work.  
    In these figures all the parameters are given with natural unit $\hbar=2m=r_0=1$.
     {\bf (b):} $\sigma_{\rm tot}(k)$ as a function of $U_0$, for the systems with $U_1=0$, $k=37$, and $\omega=10$. {\bf (c):}  $\sigma_{\rm tot}(k)$ as a function of $U_0$, for the case with $U_1=10 U_0$, $k=37$,$\omega=1$. {\bf (d):} $\sigma_{\rm tot}(k)$ as a function of $k$, for the case with  $U_0 = U_1 = 10$ and $\omega = 1$ (orange dots connected by lines, and blue squares) and $U_0 = 100$,$U_1=0$ and $\omega = 3$ (red dots connected by lines, and black squares).}
    \label{result}
\end{figure*} 

\section{Results for Shaking Spherical Square-Well Model}
\label{sw}

Now we  illustrate the generalized EA  derived above  via  
an example. Explicitly, we consider the scattering of a particle on a spherical square-well with shaking depth (Fig.~\ref{result}(a)):
\begin{eqnarray}
    U({\bm r},t)=
    \left\{
        \begin{aligned}
            U_0\cos(\omega t)+U_1, &\quad \quad r\leqslant r_0\\
            0, &\quad \quad r>r_0\\
        \end{aligned}
    \right.,
\end{eqnarray}
with $r_0$ being the width of the square well, $\omega$ being the shaking angular frequency,  and $U_{0(1)}$ being the amplitude of the  shaking (non-shaking) parts of the well depth.
In the following discussions we use the natural unit $\hbar=m=r_0=1$.

% \textcolor{blue}{
% We initially computed the variation of the scattering cross-section with respect to $\theta$, as illustrated in FIG.\ref{theta}. Within the high-energy scattering regime, a distinct peak emerges in the total scattering cross-section $\sigma_{tot}$ at $\theta=0$. With all other system parameters held constant, it's observed that as the incident kinetic energy increases, the half-width of the peak at $\theta=0$, characterized by $\Delta$, becomes narrower. Consequently, in the context of high-energy periodic scattering, we can approximate the scattering predominantly occurring at $\theta=0$.
% }
% 在FGI.\ref{result}中，我们对比了程函近似的结果和精确求解的结果，得到了很好的一致性。势能函数越小，动能越大时，程函近似的结果越好。

We calculate the total cross section$\sigma_{\rm tot}(k)$ for various parameters, with both the exact numerical approach and the generalized EA  developed in the above sections (Eqs.~(\ref{opt}) and (\ref{ff0})). In Figs.~\ref{result}(b) and (c) we show the $\sigma_{\rm tot}$
 as functions of the shaking amplitude $U_0$
 for the cases with fixed $U_{1}$, $\omega$ and $k$. Additionally, in Fig.~\ref{result}(d) we show the $\sigma_{\rm tot}$ as a function of incident momentum $k$,  
for the cases with fixed $U_{0, 1}$ and $\omega$.

Figs.~\ref{result}(b-d) show that for these parameters the results given by the generalized EA  consists very well the ones given by exact numerical calculation. These results show the applicability of the generalized EA.

Moreover,  the error of the generalized EA  is slightly increased  when  $U_0\gtrsim 15$, for the system of Fig.~\ref{result}(c) with $U_1=10U_0$. This may be explained as follows.
For this system when $U_0\gtrsim 15$ the depth $U_1$ of the static part of the square well is as large as $150$. As a result, the condition $E\gg U_\ast$ of Eq.~(\ref{con}) is not satisfied as well as in the cases with small $U_0$. Similarly, for the system of Fig.~\ref{result}(d) the error of the generalized EA  is slightly increased  in the cases with small incident momentum $k$, because in these cases the high-energy condition~(\ref{con}) is not satisfied so well.

%This may be explained as follows. 
%As shown in Eqs.~(\ref{wf}, \ref{psirt}), in the generalized EA  the scattering wave function is approximated as the the product of $e^{-iEt/\hbar}e^{ikz}$ and a slowly-varying part $\phi_{\rm EA}(\bm{r},t)$. This approximation requires that the potential energy $U({\bm r},t)$ does not changes too fast with the position ${\bm r}$. However,
%for the system of Fig.~\ref{result}(c), when $U_0\gtrsim 15$ the depth $U_1$ of the static part of the square well is as large as $150$. Thus, in the spatial region around $r=r_0$,  the absolute value of potential energy suddenly changes from zero to about $150$ or even larger. Due to such fast change, the EA does not works so well in this parameter region.

%In Fig.~\ref{result} (b-d) we compare the results obtained from the EA with those from the exact solution, and found excellent consistency. The accuracy of the perturbation approximation improves as the potential energy decreases and the kinetic energy increases.

\section{Summary}
\label{summary}
 
In this work we generalize the EA  for the Floquet scattering problems 
with 
periodical potential. We further demonstrate the generalized EA with the example of shaking spherical square-well model. The approach we developed can be used in the theoretical studies for the external-field manipulation of collisions between  particles, {\it e.g.}, the laser manipulation of collisions between atoms, nucleons or electrons.

\begin{acknowledgments}
This work is supported by the National Key Research and Development Program of China (Grant  No. 2022YFA1405300) and  the Innovation Program for Quantum Science and Technology (Grant No.~2023ZD0300700).
\end{acknowledgments}

 % \begin{widetext}
\appendix
\begin{widetext}

\section{Floquet Scattering Amplitude}
\label{fsa}

In this appendix we show why Floquet scattering amplitude can be expressed as in Eq.~(\ref{fkkp}). For a Floquet scattering problem with Hamiltonian ${\hat H}$ of Eq.~(\ref{h}), the explicit scattering state $\Psi({\bm r},t)$ 
with respect to incident momentum ${\bm k}$
satisfies the Schr\"odinger equation (\ref{se}), and can be expressed as $\Psi({\bm r},t)=e^{-{\color{black}\mathrm{i}}Et/\hbar}\psi({\bm r},t)$, with 
$E=\hbar^2k^2/(2m)$ and
$\psi({\bm r},t)=\psi({\bm r},t+T)$, as shown in Sec.~II. Furthermore, the wave function $\psi({\bm r},t)$ also satisfies the long-range outgoing boundary condition 
\begin{eqnarray}
\lim_{r\rightarrow\infty}\psi({\bm r},t)=\frac{1}{(2\pi)^{3/2}}\left[{\color{black}\mathrm{e}}^{{\color{black}\mathrm{i}}{\bm k}\cdot{\bm r}}+\sum_{n=n_\ast}^{+\infty}\frac{f_n({\hat{\bm r})}}{r} {\color{black}\mathrm{e}}^{{\color{black}\mathrm{i}}k_n r}{\color{black}\mathrm{e}}^{-{\color{black}\mathrm{i}}n\omega t} \right],
\label{a1}
\end{eqnarray}
where ${\hat {\bm r}}\equiv {\bm r}/r$ is the direction vector of ${\bm r}$, and $k_n=\sqrt{2m(E+n\hbar\omega)/\hbar}$. 
Furthermore, the Floquet scattering amplitude with respect to the incident momentum ${\bm k}$ and 
outgoing momentum $k_n{\hat {\bm r}}$ is defined as the factor $f_n({\hat{\bm r})}$ in Eq.~(\ref{a1}), {\it i.e.}, 
\begin{eqnarray}
f({\bm k}',n\leftarrow {\bm k})\equiv f_n({\hat{\bm k}}')\ {\rm of}\ {\rm Eq.~(\ref{a1})}.\label{a2}
\end{eqnarray}

Now our task is to prove the scattering amplitude defined in Eq.~(\ref{a2}) can be expressed as in Eq.~(\ref{fkkp}). For convenience, here we formally introduce the Hilbert space ${\mathscr H}_F$, which is defined as the set of all periodical functions $\eta(t)$ which satisfies $\eta(t)=\eta(t+T)$. We denote the vector of ${\mathscr H}_F$ as $|)$, and define the basis of ${\mathscr H}_F$ as:
\begin{eqnarray}
|s)\equiv e^{-{\color{black}\mathrm{i}}s\omega t},\ \ \ \ \ \ (s=0,\pm 1, \pm 2,....).\label{a3}
\end{eqnarray}
We further the inner product of two periodical functions $\eta(t)$ and $\eta'(t)$ as 
\begin{eqnarray}
(\eta'|\eta)=\frac 1T\int_0^{T} \eta'(t)^\ast \eta(t){\color{black}\mathrm{d}}t,
\end{eqnarray}
which yields that ${\rm Span}\{|0), |\pm 1), |\pm 2),...\}$ is a group of orthogonal basis of ${\mathscr H}_F$.

Now we consider the wave function $\psi({\bm r},t)$ of our problem. Due to the periodical condition $\psi({\bm r},t)=\psi({\bm r},t+T)$, $\psi({\bm r},t)$ is a ${\bm r}$-dependent element of ${\mathscr H}_F$, and can be denoted as $|\psi[{\bm r}])$. Explicitly, we have the notation correspondence: 
\begin{eqnarray}
\psi({\bm r},t)=\sum_{s=-\infty}^{+\infty}\psi_s({\bm r}){\color{black}\mathrm{e}}^{-{\color{black}\mathrm{i}}s\omega t}\ \Longleftrightarrow\ 
|\psi[{\bm r}])=\sum_{s=-\infty}^{+\infty}\psi_s({\bm r})|s).
\end{eqnarray}
Moreover, with the new notation the equation satisfied by $\psi({\bm r},t)$ can be re-expressed as
\begin{eqnarray}
{\hat {\cal H}}|\psi[{\bm r}])=E|\psi[{\bm r}]),
\end{eqnarray}
with
\begin{eqnarray}
{\hat {\cal H}}=-\frac{\hbar^2}{2m}\nabla^2\otimes{\hat {\cal I}}+\sum_{s=-\infty}^{+\infty}U_s({\bm r}){\hat{\cal C}}^{ s}.\label{nse}
\end{eqnarray}
Here ${\hat {\cal I}}$ is the unit operator of ${\mathscr H}_F$, and ${\hat{\cal C}}=\sum_{n=-\infty}|n+1)(n|$. In addition, the functions $U_s({\bm r})$ ($s=0,\pm 1, \pm 2,...$) are related to the potential $U({\bm r},t)$ via 
\begin{eqnarray}
U({\bm r},t)=\sum_{s=-\infty}^{+\infty}U_s({\bm r}){\color{black}\mathrm{e}}^{-{\color{black}\mathrm{i}}s\omega t}.
\end{eqnarray}
Furthermore, the long-range boundary condition (\ref{a1}) of $\psi({\bm r},t)$ can be re-expressed as 
\begin{eqnarray}
\lim_{r\rightarrow\infty}|\psi[{\bm r}])=\frac{1}{(2\pi)^{3/2}}\left[{\color{black}\mathrm{e}}^{{\color{black}\mathrm{i}}{\bm k}\cdot{\bm r}}|0)+\sum_{n=n_\ast}^{+\infty}\frac{f_n({\hat{\bm r})}}{r} {\color{black}\mathrm{e}}^{{\color{black}\mathrm{i}}k_n r}|n) \right].
\label{a8}
\end{eqnarray}

%On the other hand, according to the Floquet theory, our scattering problem defined in the Hilbert space ${\mathscr H}$ of the spatial motion of the particle
% can be mapped to the one in the Hilbert space ${\mathscr H}\otimes{\mathscr H}_F$, with 
%\begin{eqnarray}
%{\mathscr H}_F={\rm Span}\{|0), |\pm 1), |\pm 2),...\}
%\end{eqnarray}
%being an auxiliary Hilbert space, and the basis satisfies . In the following we show the detail of this mapping and prove that the scattering amplitude defined in Eq.~(\ref{a1}) can be expressed as in Eq.~(\ref{fkkp}). 

%In summary, we can derive the wave function  $\psi({\bm r},t)$, which is is related to the scattering wave function $\Psi(\bm r,t)$ via Eq.~(\ref{wf}), with the following  steps: (1) Derive $|\psi[{\bm r}])$ by
%solving the Schr\"odinger equation (\ref{nse})  with the outgoing boundary condition (\ref{a8}).
%(2) Expand  $|\psi[{\bm r}])$ as $|\psi[{\bm r}])=\sum_{s=-\infty}^{+\infty}\psi_s({\bm r})|s)$. 
%(3) Then the wave function  $\psi({\bm r},t)$ is just given by $\psi({\bm r},t)=\sum_{s=-\infty}^{+\infty}\psi_s({\bm r})e^{is\omega t}$.

The above results yield that
the term $f_n({\hat{\bm r})}$, which is originally defined in Eq.~(\ref{a1}), is also the scattering amplitude of the latter scattering problem  with Hamiltonian  {\it time-independent} ${\hat {\cal H}}$ and incident wave function $\frac{{\color{black}\mathrm{e}}^{{\color{black}\mathrm{i}}{\bm k}\cdot{\bm r}}}{(2\pi)^{3/2}}|0)$. 
Notice that this scattering problem is defined in the product space ${\mathscr H}\otimes{\mathscr H}_F$. 
Since ${\hat {\cal H}}$ is time-independent, we can use the standard scattering theory to analyze the properties of the scattering amplitude.  Using Eq.~(\ref{a2}) and the scattering theory, we finally have
\begin{eqnarray}
f({\bm k}',n\leftarrow {\bm k})
\equiv f_n({\hat{\bm k}'})=-(2\pi)^{1/2}m\int {\color{black}\mathrm{d}}{\bm r}{\color{black}\mathrm{e}}^{-{\color{black}\mathrm{i}}k_n{\hat {\bm k}'}\cdot{\bm r}}(n|\left[\sum_{s=-\infty}^{+\infty}U_s({\bm r}){\hat{\cal C}}^{ s}\right]|\psi[{\bm r}]).
\label{a10}
\end{eqnarray}
Using Eqs.~(\ref{a3}-\ref{a8}), we  directly find that Eq.~(\ref{a10}) is just Eq.~(\ref{fkkp}).

%The above results yield that the scattering problem with periodical Hamiltonian $H$ and incident wave function $\frac{e^{i{\bm k}\cdot{\bm r}}}{(2\pi)^{3/2}}$, which is defined in the Hilbert space ${\mathscr H}$ of the spatial motion of the particle, can be mapped to a scattering problem with Hamiltonian ${\hat {\cal H}}$, which is defined in the product space ${\mathscr H}\otimes{\mathscr H}_F$ and incident wave function $\frac{e^{i{\bm k}\cdot{\bm r}}}{(2\pi)^{3/2}}|0)$. In particular, we have proved that

%As a result, we can derive the scattering amplitude $f_n({\hat{\bm r})}$ by solving the Schr\"odinger equation (\ref{nse})  with the outgoing boundary condition (\ref{a8}).

%For the convenience of our discussion, let us first define some symbols. We denote the states in ${\mathscr H}$, ${\mathscr H}_F$ and ${\mathscr H}\otimes{\mathscr H}_F$ as $|\rangle$, $|)$ and $|\rangle\!\rangle$, respectively. 
%As before, we describe each state $|\eta\rangle\in {\mathscr H}$  via the wave function $\eta({\bf r})\equiv\langle{\bm r}|\eta\rangle$, with $|{\bm r}\rangle$ being the eigen-state of atomic position. In this representation, a state $|\eta\rangle\!\rangle$ in ${\mathscr H}\otimes{\mathscr H}_F$ can also be described by its wave function 
%$\langle{\bm r}|\eta\rangle\!\rangle\equiv |\eta[{\bm r}])$.

\section{Derivations of Eqs.~(\ref{fea2}) and (\ref{ftheta})}
\label{pro1}

In this appendix we derive Eqs.~(\ref{fea2}) and (\ref{ftheta}) of the maintext. We first notice that, under the condition (\ref{ccon}),
${\bm k}'-{\bm k}$ is approximately perpendicular to the direction of ${\bm k}$ ({\it i.e.}, the $z$-direction), and thus $({\bm k'}-{\bm k})\cdot {\bm r}\approx({\bm k'}-{\bm k})\cdot {\bm b}$. 
Substituting this result and the conditions $n=0$ into Eq.~\eqref{fea}, we obtain
\begin{eqnarray}
    f({\bm k}',0\leftarrow{\bm k})\approx-\frac{m\omega}{(2\pi)^{2}}
\int_{0}^{T} \!\!\!{\color{black}\mathrm{d}}t 
    \!\int_{-\infty}^{+\infty} \!\!\! {\color{black}\mathrm{d}}\bm b \int_{-\infty}^{+\infty}{\color{black}\mathrm{d}}z \left\{{\color{black}\mathrm{e}}^{-{\color{black}\mathrm{i}}({\bm k'}-{\bm k})\cdot {\bm b}} 
    U(\bm r,t)\phi_{\rm EA}(\bm r,t)\right\}.\label{fkpka}
    \end{eqnarray}
    Furthermore, Eq.~(\ref{appe}) yields that 
\begin{eqnarray}
   U({\bm r},t)\phi_{\rm EA}({\bm r},t)=  {\color{black}\mathrm{i}}\hbar\frac{\partial }{\partial t}\phi_{\rm EA}({\bm r},t)+{\color{black}\mathrm{i}}\hbar v_z \frac{\partial }{\partial z}\phi_{\rm EA}({\bm r},t).\label{up}
   \end{eqnarray}
   Substituting Eq.~(\ref{up}) into Eq.~(\ref{fkpka}), and using  the property (\ref{ppc}) of $\phi_{\rm EA}({\bm r},t)$, as well as the facts
 \begin{eqnarray}
 \phi_{\rm EA}({\bm r},t)\bigg\vert_{z=-\infty}&=&1;\\
  \phi_{\rm EA}({\bm r},t)\bigg\vert_{z=+\infty}&=&\exp\left\{ -{\color{black}\mathrm{i}}\frac{1}{v_z\hbar}\int_{-\infty} ^{+\infty}
    U\bigg[
    {\bm b}, z', {\tilde t}(t,z,z')
    \bigg]{\color{black}\mathrm{d}}z'
         \right\},
   \end{eqnarray}  
 which are given by the expression (\ref{cr}) of $\phi_{\rm EA}({\bm r},t)$ and $mv_z=\hbar k$, we immediately obtain Eq.~(\ref{fea2}).

Furthermore, when the potential $U({\bm b},z,t)= U(b,z,t)$ is independent of the direction of ${\bm b}$. In this case $f({\bm k}'-{\bm k})$ only depends on the norm $k=|{\bm k}|=|{\bm k}'|$ and the angle $\theta$ between ${\bm k}$ and ${\bm k}'$. Without loss of generality, we take the direction of $\bm k'-\bm k$ to be along the $x$-axis, and define $\phi$ to be the angle between $\bm b$ and the $x$-axis. Since we consider the cases with small $\theta$, we have
\begin{equation}
    (\bm k'-\bm k)\cdot \bm b\approx k \theta  b \cos \phi.\label{kkpa}
\end{equation}
% 在小角度近似下，$\phi$可以看做是方位角，即$\vec{b}$与$y$轴的夹角，对于方位角$\phi$对称的势，
Substituting Eq.~(\ref{kkpa}) into Eq.~(\ref{fea2}), and using the facts $\int\bm b=\int_0^{+\infty}b{\color{black}\mathrm{d}}b\int_{0}^{2\pi} {\color{black}\mathrm{d}}\phi$, and $\int_{0}^{2\pi} {\color{black}\mathrm{e}}^{-{\color{black}\mathrm{i}} \alpha \cos\phi}\,{\color{black}\mathrm{d}} \phi = 2\pi J_0(\alpha)$, 
we can obtain Eq.~(\ref{ftheta}).

\end{widetext}

\bibliography{references}

\begin{thebibliography}{40}
\bibitem [{2pa()}]{2page}%
  \BibitemOpen
  \href@noop {} {}\bibinfo {note} {{\color{black}Notice that the scattering wave function $\Psi({\bm r},t)$ given by EA is different from the one given by the semi-classical approximation (SCA)\cite{weinberg2015lectures}, which is the three-dimensional generalization of the Wentzel-Kramers-Brillouin (WKB) approximation. This difference can be explained as follows. For simplicity, we consider the case with a time-independent potential, and denote the scattering wave function given by the SCA as $\Psi_{\rm SCA}({\bm r})$. As shown in Ref. \cite{weinberg2015lectures}, the phase of $\Psi_{\rm SCA}({\bm r})$ is determined by the action of a classical trajectory with respect to ${\bm r}$. Thus, for each ${\bm r}$, to determine the value of $\Psi_{\rm SCA}({\bm r})$, one should solve an individual classical Hamiltonian equation, and then integrating the corresponding Lagrangian with time. In contrast, as shown in our maintext, under the EA, one can derive the phase of the wave function ({\it i.e.}, $\mathrm{i}\frac{1}{v_z\hbar}\int_{-\infty} ^{z} U[ {\bm b}, z']\mathrm{d}z$ for the case with time-independent potential) by directly integrating the potential energy, without solving any classical dynamical equation. Additionally, we should also notice that the conditions of the WKB approximation and EA are different. Explicitly, the condition for the former can be expressed as $k\gg 2\pi/l_U$ with our notations, while the one for EA is Eq.~(\ref{con}), {\it i.e.}, both $k\gg 2\pi/l_U$ and $E\gg U_\ast$ }}\BibitemShut {NoStop}%
\end{thebibliography}
\end{document}